\documentclass[sigplan,10pt,screen]{acmart}

\def\Title{A Concurrency-Agnostic Protocol for Multi-Paradigm Concurrent Debugging Tools}
\def\ShortTitle{A Concurrency-Agnostic Protocol for Concurrent Debugging Tools}

\def\PdfTitle{\Title}
\def\PdfSubject{Debugging for complex concurrent systems that combine various concurrency models.}
\def\PdfAuthors{{Stefan Marr, Carmen Torres Lopez, Dominik Aumayr, Elisa Gonzalez Boix, Hanspeter Mössenböck}}
\def\PdfKeywords{Debugging, Tooling, Concurrency, Breakpoints, Stepping, Visualization}

\usepackage[utf8]{inputenc}
\usepackage{textgreek}
\usepackage{fixltx2e} 
\usepackage{ifpdf}
\usepackage{graphicx}

\newif\ifhtml
\ifpdf
\htmlfalse
\else
\htmltrue
\fi

\usepackage{scripts/collab}  

\ifpdf
    \PassOptionsToPackage{bookmarks=true,bookmarksnumbered=true,
        pdfpagemode={UseOutlines},plainpages=false,pdfpagelabels=true,
        colorlinks=true,linkcolor={black},citecolor={black},urlcolor={black},
        pdftitle={\PdfTitle},%
        pdfsubject={\PdfSubject},%
        pdfauthor={\PdfAuthors},%
        pdfkeywords={\PdfKeywords},
        pdftex,unicode}{hyperref}

\else
    \ifhtml
    \else
        \PassOptionsToPackage[%
           colorlinks=true,%
           linkcolor={black},%
           citecolor={black},%
           urlcolor={black}]{hyperref}
    \fi
\fi

\newcommand{\ie}{i.e.\xspace}
\newcommand{\eg}{e.g.\xspace}

\usepackage{letltxmacro}
\LetLtxMacro{\OrgCitep}{\citep}
\renewcommand{\citep}[1]{\,\OrgCitep{#1}}

\newcommand{\code}[1]{\lstinline!#1!}
\usepackage{listings}
\usepackage{ifthen}
\newcommand{\citeurl}[5]{%
#1\footnote{\emph{#2}%
          \ifthenelse{\equal{#3}{}}
                     {}
                     {, #3}
          \ifthenelse{\equal{#4}{}}
                     {}
                     {, access date: #4}
          , \url{#5}}}

\newcommand{\circled}[1]{\textcircled{\scriptsize\textsf{#1}}}

\usepackage[nameinlink]{cleveref}
\Crefformat{lstlisting}{#2Lst.\,#1#3}
\crefformat{lstlisting}{#2lst.\,#1#3}

\ifhtml
  \def\SOMns{SOM\HCode{<span style="font-variant: small-caps;">}ns\HCode{</span>}\xspace}
\else
  \def\SOMns{SOM{\sc ns}\xspace}
\fi
\def\Kompos{K\'{o}mpos\xspace}
\def\NumBreakpoints{21\xspace}
\def\NumSteppingOps{20\xspace}

\collabAuthor{sm}{red}{Stefan}
\collabAuthor{egb}{blue}{Elisa}
\collabAuthor{ctl}{orange}{Carmen}
\collabAuthor{hm}{violet}{Hanspeter}
\collabAuthor{dau}{violet}{Dominik}

\usepackage{booktabs}   

\setcopyright{acmlicensed}
\acmPrice{15.00}
\acmDOI{10.1145/3133841.3133842}
\acmYear{2017}
\copyrightyear{2017}
\acmISBN{978-1-4503-5526-1/17/10}
\acmConference[DLS'17]{13th ACM SIGPLAN International Symposium on Dynamic Languages}{October 24, 2017}{Vancouver, Canada}
\startPage{1}

\begin{document}

\title[\ShortTitle]{\Title}         


\author{Stefan Marr}
\orcid{0000-0001-9059-5180}
\affiliation{%
  \institution{Johannes Kepler University}
  \city{Linz}
  \country{Austria}}
\email{stefan.marr@jku.at}
\author{Carmen Torres Lopez}
\affiliation{%
  \institution{Vrije Universiteit Brussel}
  \city{Brussels}
  \country{Belgium}}
\email{ctorresl@vub.be}
\author{Dominik Aumayr}
\affiliation{%
  \institution{Johannes Kepler University}
  \city{Linz}
  \country{Austria}}
\email{dominik.aumayr@jku.at}
\author{Elisa Gonzalez Boix}
\affiliation{%
  \institution{Vrije Universiteit Brussel}
  \city{Brussels}
  \country{Belgium}}
\email{egonzale@vub.be}
\author{Hanspeter Mössenböck}
\affiliation{%
  \institution{Johannes Kepler University}
  \city{Linz}
  \country{Austria}}
\email{hanspeter.moessenboeck@jku.at}


\begin{abstract}

Today's complex software systems combine high-level concurrency models.
Each model is used to solve a specific set of problems.
Unfortunately, debuggers support only the low-level notions of threads and shared memory, forcing developers to reason about these notions instead of the high-level concurrency models they chose.

This paper proposes a concurrency-agnostic debugger protocol that decouples the debugger from the concurrency models employed by the target application.
As a result, the underlying language runtime can define custom breakpoints, stepping operations, and execution events for each concurrency model it supports, and a debugger can expose them without having to be specifically adapted.

We evaluated the generality of the protocol by applying it to \SOMns, a Newspeak implementation, which supports a diversity of concurrency models including communicating sequential processes, communicating event loops, threads and locks, fork/join parallelism, and software transactional memory.
We implemented \NumBreakpoints breakpoints and \NumSteppingOps stepping operations for these concurrency models.
For none of these, the debugger needed to be changed.
Furthermore, we visualize all concurrent interactions independently of a specific concurrency model.
To show that tooling for a specific concurrency model is possible, we visualize actor turns and message sends separately.


\end{abstract}



\maketitle
\renewcommand{\shortauthors}{Marr et al.}

\section{Introduction}

Building and maintaining complex concurrent systems is a hard task.
Some developers combine different high-level models to solve problems with a suitable tool\citep{CastagnaECOOP2013}.
Unfortunately, debugging support for combined concurrency models is minimal, which makes building and maintaining complex concurrent systems even harder.

For more than three decades, debugging support for concurrency models has been studied for each model in isolation\citep{McDowell:1989}.
As a result, thread-based languages such as Java and C/C++ have debuggers aware of threads and locks.
Similarly, ScalaIDE and Erlang provide support for actors and message sending.

However, support for debugging combined concurrency models is still missing.
The main challenge is to identify a common representation for concurrency models so that a debugger does not need specialized support for each model.
For example, there are four main interpretations of the actor model, each of which has been implemented in different variations\citep{DeKoster:2016:YAT}.
For comprehensive debugging support of all these variations, a debugger needs to abstract from the concrete concurrency model and provide a common set of abstractions instead.
This would allow us to use the same debugger without changes for the different concurrency models and their numerous variations.

This paper presents the \Kompos protocol, a \emph{concurrency-agnostic protocol} to enable debuggers to support a wide range of concurrency models.
Using the \Kompos protocol, we built the \Kompos debugger for online debugging of complex concurrent systems that combine \emph{communicating event loops} (CEL)\citep{ELangActors}, \emph{communicating sequential processes} (CSP)\citep{CSP}, \emph{software transactional memory} (STM)\citep{Harris:2005}, \emph{fork/join}\citep{Cilk}, and shared-memory \emph{threads and locks}.
Based on the concurrency-agnostic protocol, \Kompos supports a rich set of breakpoints for the various concurrency abstractions, a rich set of stepping semantics to explore program behavior, a generic visualization of interactions between concurrent entities, as well as an actor-specific visualization of turns and message sends.
This evaluation shows that the \Kompos protocol is (1) general enough to support advanced debugger features for shared-memory and message-passing models, and (2) that it supports tools using the provided data independently of any concurrency model, while preserving the ability to build tools specific to a single model.

\Kompos is a debugger for \SOMns, a Newspeak implementation\citep{Bracha:10:NS} based on Truffle\citep{Wurthinger:2012:SelfOptAST}.
\SOMns' debugger support is built on Truffle's tooling and debugger features\citep{VanDeVanter:2015:BDO,Seaton:2014:DFS}.
\SOMns supports the five aforementioned concurrency models and implements breakpoints, stepping, and a tracing mechanism for them.
Because of the concurrency-agnostic design of the \Kompos protocol, the \Kompos debugger is independent from these concurrency models.

The contributions of this paper are:

\begin{itemize}
  \item An analysis of the major shared-memory and message passing concurrency models to identify abstractions for a concurrency-agnostic debugger protocol.
  \item A concurrency-agnostic debugger protocol that enables custom breakpoints, stepping, and visualization.
  \item An implementation of the \Kompos protocol as part of the \Kompos debugger and \SOMns.
  \item An evaluation of the protocol based on CEL, CSP, STM, fork/join, and threads and locks.
\end{itemize}



\section{Background}
\label{sec:background}

This section discusses existing debugger protocols and introduces the Truffle debugger, on which we build our work. 

\subsection{Debugger Protocols}
\label{sec:dbg-protocol}

Runtime systems and integrated development environments (IDE) typically communicate via a \emph{debugger protocol}.
This includes the \citeurl{Java Debug Wire Protocol (JDWP),}{Java Debug Wire Protocol}{Oracle Inc.}{2017-05-16}{https://docs.oracle.com/javase/8/docs/technotes/guides/jpda/jdwp-spec.html}
the \citeurl{GDB machine interface,}{GDB/MI Interface}{The GNU Project Debugger}{2017-05-16}{https://sourceware.org/gdb/onlinedocs/gdb/GDB_002fMI.html}
the \citeurl{Chrome DevTools protocol,}{Chrome DevTools Protocol}{Google}{2017-05-16}{https://chromedevtools.github.io/devtools-protocol/}
and the \citeurl{Visual Studio Code debug protocol.}{VS Code Debug Protocol}{Microsoft}{2017-05-16}{https://github.com/Microsoft/vscode-debugadapter-node/tree/master/protocol}

%
%

These protocols define commands to interact with the program, to request information about threads, stack frames, local variables, objects, memory.
They also communicate events, \eg, when a breakpoint was hit.
While the protocols differ in format and structure, they cover a similar set of common debugger features.
For instance, they allow a user to define a breakpoint for a specific source location, possibly with filters and conditions attached to it.

The use of a debugger protocol also decouples runtime systems and debuggers facilitating the construction of new tools.
However, these protocols are often designed for sequential or threaded languages.
As such, their support for debugging concurrency abstractions is rather limited.
For instance, JDWP and GDB can show a list of running threads or control execution of a particular thread.
GDB also has support to request information about Ada tasks.
In Chrome, a step-into-async operation is the only explicit support for concurrent stepping.
In short, the protocols are limited to these specific concurrency concepts. 

Thus, custom breakpoints or stepping operations for concurrency models are not supported.
Instead, the protocols support a fixed set of breakpoint and stepping operations.
Any extension
requires changing the protocol.

In this paper, we argue that a protocol similar to the ones mentioned above forms a foundation for the basic debugger features, \ie, to interact with the program and request basic information such as values or local variables and objects.
However, in contrast to the classical debugger protocols, we propose a \emph{concurrency-agnostic protocol} that supports a wide range of concurrency models.
Our implementation in \SOMns is inspired by the Visual Studio Code protocol (cf. \cref{sec:impl}), but as detailed later, the \Kompos protocol abstracts completely from different breakpoints and stepping operations and thereby provides the necessary flexibility to support custom semantics for different concurrency models.

\subsection{Truffle Debugger: A Language-Agnostic Debugging Framework}
\label{sec:bg-truffle}
\SOMns is built on top of Truffle, a framework for AST-based interpreters\citep{Wurthinger:2012:SelfOptAST}.
Part of the framework is support for interpreter instrumentation, which is used, \eg, for language-agnostic debugging and execution monitoring.
The framework provides Truffle-languages with a classic breakpoint-based debugger for sequential code\citep{Seaton:2014:DFS}, which we use in \SOMns for the basic sequential stepping and breakpoint support.

One key element of the framework is its use of tags for the AST nodes\citep{MVDV:2017:MoreVMs}.
For the Truffle debugger support, a language annotates AST nodes with tags for \code{Statement}, \code{Call}, and \code{Root}.
Based on these tags, the debugger determines the target nodes for line breakpoints, single stepping, step over, and returning from a method.
\SOMns uses the same mechanism for AST node tags to encode additional information, which is used to recognize concurrency-related operations as well as generic syntax information such as keywords.
The \Kompos debugger can use this information for debugging and syntax highlighting.

\section{Which Concurrency Concepts are Relevant for Debugging?}
\label{sec:concepts}


This section analyzes concurrency models to identify the basic concepts that need to be supported by a debugger protocol to enable concurrency-related breakpoint and stepping operations.
As a basic categorization, we distinguish shared-memory and message-passing models\citep{Almasi:1989:HPC}.
We analyze instances for both types of models
including threads and locks, STM, and fork/join parallelism as shared-memory models, and CEL and CSP as message-passing models.
To account for more advanced debugger features, we also analyze what information would be needed to visualize these concepts.

\subsection{Threads and Locks (T\&L)}

Shared-memory models, as supported by e.g., C/C++, Java, C\#, have a wide range of concepts for simultaneous execution and access control to shared resources.
For brevity, this analysis includes only threads, locks, condition variables, and object monitors.
Other constructs are left to future work.

Threads are the active entities that execute code.
Locks and object monitors enable the synchronization of threads, i.e., they restrict thread interactions on shared resources to enforce correctness properties.
Condition variables are used to communicate between threads that certain conditions have changed.
%
Furthermore, object monitors, as known from Ada or Java, are a structured synchronization mechanism that uses a lock to protect some shared resource in the dynamic scope of some object method or code block.
In contrast, locks and condition variables are entities with which a thread can interact in unstructured ways.

For debugging, it needs to be possible to step through the execution of a thread and set breakpoints on statements.
This should include the standard operations to step into or over a call to, and return from a method.
However, the debugger should also provide stepping and breakpoints for concurrency abstractions such as locks, object monitors, and conditions variables. 
This would allow developers to check for incorrect synchronization.
For locks, the debugger should be able to step from an \code{acquire} operation to the corresponding \code{release} operation to see how they relate.
That also helps to detect unbalanced acquire/release operations which, e.g, can lead to starvation if locks are not released.
Similarly, for object monitors, the debugger should allow developers to set a breakpoint on entering and exiting the monitor.
For condition variables, stepping between the \code{wait} and \code{notify} operations allows developers to observe their effects on, and communication with other threads.
From the point where a thread is created in a program, the debugger should allow developers to set a breakpoint on its execution, or to step into its execution.
Similarly, when a thread terminates, the debugger should allow developers to suspend execution of the thread that joins with it and waits for termination.

The high-level interactions of threads entering monitors and using locks or condition variables are also relevant for visualization, which is useful for
identifying unintended interactions or missing synchronization.

\subsection{Communicating Event Loop Model (CEL)}

The CEL model is a variation of the actor model, used by languages such as E\citep{ELangActors} and AmbientTalk\citep{VanCutsem2014112}.
The main concepts are actors, messages, promises, and turns.
Actors are the active entities, each with its own event loop.
Actors execute messages as \emph{turns}, \ie, one by one, in the order they are received.
They contain a set of objects and interact via asynchronous messages because actors do not share memory.
Promises are 
\emph{eventual values}.
They can establish a data dependency between actors, \eg, as placeholders for the return value of an asynchronous message.
Since the CEL model includes only non-blocking abstractions, promise values can be accessed only via callbacks, which are executed as turns on an actor.

When debugging such CEL actors, developers should be able to step through turns of a specific actor to see which messages are received.
This means, the debugger should combine normal sequential stepping within a turn with the ability to skip sequential operations and step to the next turn.
When sending messages, suspending the actor's execution before a message is sent would allow developers to inspect its parameters and target object.
Similarly, following the execution to observe how promises are resolved and how callbacks on them are executed after resolution can help developers to identify communication issues or unexpected values.
For all these operations, the debugger should be able to either explicitly step through them or define breakpoints.

For a visualization, the high-level interactions of actors with messages and promises are relevant. For example, a visualization of the order in which turns, \ie, messages are executed could help to identify synchronization issues and bad message interleavings.

\subsection{Communicating Sequential Processes (CSP)}


The main concepts in CSP are processes, channels, and messages.
Processes are the active entities that execute code.
Channels are first-class elements that connect processes and allow them to communicate by passing messages with rendezvous semantics.
A message is a specific datum exchanged via a channel. 
Like CEL, CSP is a message passing model. 
However, CSP uses blocking semantics and has no notion of turns, which makes it very different to CEL. 

For debugging, it should be possible to follow the sequential execution of a process, from its creation to the end, like in threads.
But, similar to CEL, the debugger should be able to step through message sends, \ie, in this case channel operations, to identify the communication partners and their state, or put breakpoints on these operations.
Furthermore, it should account for the rendezvous semantics of channels, to step from a receiver to the continuation in the sender.
This is useful since channels can be passed around, which might lead to the wrong processes communicating with each other.

For visualization, the communication between processes and the run-time network of channels is relevant, \eg, to identify communication partners or lack-of-progress issues.

\subsection{Software Transactional Memory (STM)}

STM provides a wide range of notions for transactions, \eg, open or closed, optimistic or pessimistic. 
For brevity, this analysis includes only the basic concepts of threads and transactions.
Threads are the same as for other shared-memory models.
Transactions, on the other hand, introduce a dynamic scope, in which all modifications to shared state are applied either atomically or not at all.

The debugger should be able to interact with transactions in a way that makes it possible to observe the behavior when transactional conflicts occurs.
Thus, developers should be able to set breakpoints or step through the execution of transactions to the final commit operation.
Being able to stop right before a transaction commits allows developers to examine transaction interactions.
The developer should also be able to step between transactions to follow the high-level flow of program elements interacting on shared state.

For a visualization, the transactions executed on threads and their ordering could help to identify missing synchronization, unintended dependencies, or performance issues.

\subsection{Fork/Join Parallelism (F/J)}

Fork/Join programming enables parallel divide-and-conquer algorithms.
The main abstraction is an asynchronously executing task, which produces a result eventually.
The model is only concerned with decomposing problems into a structure of tasks that synchronize based on fork and join operations.
Other forms of synchronization are left out of the model.

A debugger should focus on these fork and join operations.
Breakpoints and stepping should enable developers to explore the recursive structure of spawns and joins.
Visualizing these spawn and join dependencies may help to understand the recursive nature of complex fork/join programs.

\subsection{Analysis and Conclusion}
\label{sec:req}

\begin{table*}
\caption{A Taxonomy of Concurrency Concepts Relevant for Debugging.}
\label{tab:conc-concepts}
\begin{center}
\footnotesize
\begin{tabular}{l l l l l l}
\toprule
                 & T\&L               & CEL             & CSP             & STM         & F/J       \\
Activities       & threads            & actors          & processes       & threads     & tasks      \\[4pt]
Dynamic Scopes   & object monitors    & turns           &                 & transactions &           \\[4pt]
Passive Entities & conditions         & messages        & channels        &             &           \\
                 & locks              & promises        & messages        &             &           \\[4pt]
Send Operations  & acquire lock       & send message    & send message    &             &           \\
                 & signal condition   & resolve promise &                 &             &           \\[4pt]
Receive Operations& release lock      &                 & receive message & join thread & join task \\
                 & wait for condition &                 & join process    &             &           \\
                 & join thread        &                 &                 &             &           \\
\bottomrule
\end{tabular}
\end{center}
\end{table*}

The above discussion identified the main concepts for CSP, CEL, F/J, STM, and T\&L.
This section categorizes them to establish abstractions for a debugger protocol. 

\textbf{Activities} are the active entities executing code.
This includes threads, actors, processes, and fork/join tasks.
\textbf{Dynamic scopes} are well-structured and nested parts of a program's execution during which certain concurrency-related properties hold, \eg, during a transaction, while executing code under an object monitor, or during an actor turn.
\textbf{Passive entities} are entities that do not act themselves, but are acted upon.
For example, we consider messages and promises as passive entities of the CEL model, while the ones of CSP are channels and messages.
%
Note that we do not consider normal objects as passive entities, because it was not needed.

To model interactions between these entities, we use send and receive operations.
A \textbf{Send Operation} is an interaction that initiates communication or synchronization.
A \textbf{Receive Operation}, is an interaction that reacts to a communication or synchronization operation.
Consequently, we consider acquiring a lock or signaling a condition to be send operations and joining with a thread is a receive operation.

\Cref{tab:conc-concepts} gives an overview of the identified categories per concurrency model.
These categories of entities and operations are used as foundation for the \Kompos protocol and detailed in the following section.

\section{A Concurrency-Agnostic Debugger Protocol}
\label{sec:protocol}

\begin{figure}
\centering
\ifpdf
\includegraphics{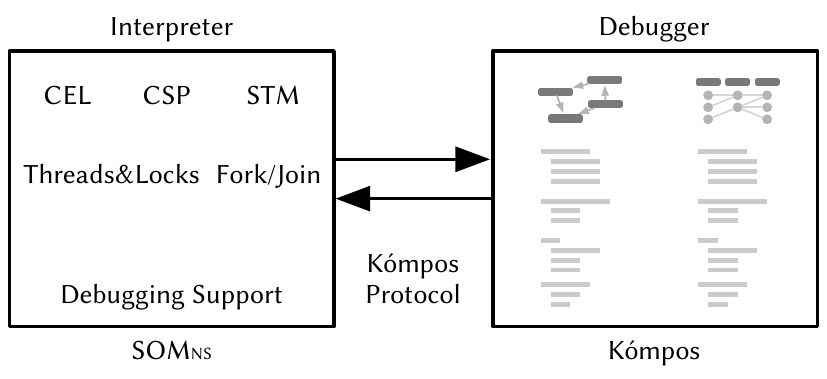}
\else
\HCode{<img src="figures/protocol-overview.svg" />}
\fi
\caption{General architecture: Interpreter and debugger communicate via the concurrency-agnostic \Kompos protocol. The interpreter provides the implementation of the different concurrency models and debugging support.}
\label{fig:architecture}
\vspace{-1em}
\end{figure}

To build a concurrency-agnostic debugger, we devise a protocol for communication between the debugger and interpreter that can support the breakpoints, stepping operations, and visualizations envisioned in \cref{sec:concepts}, without merely enumerating the concurrency concepts.
The goal is that only the language implementation knows the specifics for each concurrency model, while the debugger remains independent of them.
\Cref{fig:architecture} shows the architecture of such a system.
An interpreter with support for various concurrency models is connected to a debugger via the concurrency-agnostic \Kompos protocol.

\subsection{High-Level Overview of the Protocol Concepts}


From \cref{sec:req} we derive that concurrency concepts relevant for breakpoints, stepping operations, and visualization can be modeled based on \emph{activities}, \emph{dynamic scopes}, \emph{passive entities}, \emph{send operations}, and \emph{receive operations}. 
Using these basic notions, a protocol is independent of any specific concurrency model. 
However, the debugger requires meta data to match debugging operations and concrete concurrency concepts.
Thus, when the debugger connects to the interpreter, it receives meta data with details on the supported concurrency models, the entities they define, their breakpoints, and their stepping operations.
This information is mostly opaque to the debugger.
For breakpoints and stepping, it is sufficient to match opaque identifiers (cf. \cref{sec:protocol-details}).

\paragraph{Concurrency Concepts.}

As discussed in \cref{sec:req}, \emph{activities} are active entities that execute code.
The debugger protocol uses this notion, \eg, to offer stepping operations that are specific to an activity type.
\emph{Dynamic scopes} are used, for instance to determine possible stepping operations.
Some stepping operations are only available during a transaction or while holding an object monitor.

\emph{Passive entities}, \emph{send}, and \emph{receive operations} are used for the visualization of concurrent interactions.
However, they are currently not used in the context of pausing/resuming program execution or performing step-by-step execution.

\paragraph{Debugger Concepts.}

For debugging, the protocol includes various other concepts.
For brevity, we discuss only the ones distinct from other debugger protocols (cf. \cref{sec:dbg-protocol}).

A \emph{source} is either a file or some other form of source text.
The source text has to be annotated with \emph{source tags} to identify the semantic elements contained in a source range.
For instance, tags can indicate the source locations of message sends or lock operations so that the debugger can show breakpoints or stepping operations.
As mentioned in \cref{sec:bg-truffle}, \SOMns employs Truffle's tagging mechanism to annotate AST nodes in the interpreter.
The \Kompos protocol is used to send this information to the debugger.

A \emph{breakpoint} type defines suspension points, which may be related to concurrency concepts.
For example, one breakpoint type could be for the point where a message is received by an actor.
They are distinguished by name and define which source tags they apply to.
Thus, the debugger does not need to know the relationship between tags and breakpoints.
Instead, it can treat tags as opaque identifiers.

A \emph{stepping operation} type defines an operation to follow program execution sequentially or concurrently.
Similar to breakpoint types, stepping operation types are distinguished by name.
Furthermore, they define criteria to determine whether the operation is applicable in the current dynamic context.
Such applicability criteria include source tags, the activity type executing the currently suspended code, as well as the dynamic scopes active for the current execution.

\subsection{Example: Breakpoints and Stepping for an Atomic Block}
\label{sec:protocol-example}

\begin{figure}
\centering
\ifpdf
\includegraphics{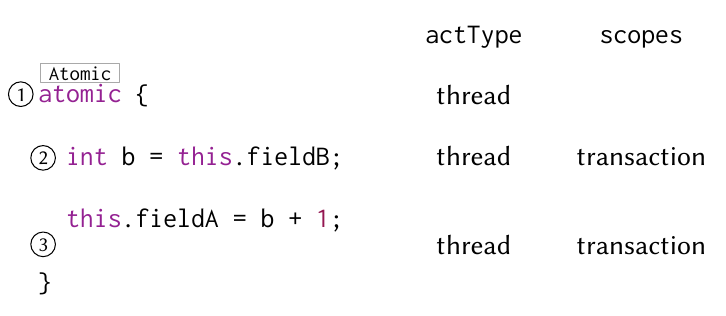}
\else
\HCode{<img src="figures/stm-example.svg" />}
\fi
\caption{Example program using an atomic transaction.
The debugger recognizes the \code{atomic} keyword via the \code{Atomic} tag.
When execution suspends at one of the three program points indicated with a number, the debugger receives location, indicated activity type, and active dynamic scopes.}
\label{fig:stm-example}
\vspace{-1em}
\end{figure}

This section discusses an example to illustrate the protocol.
Consider the program fragment in \cref{fig:stm-example}  that uses an \code{atomic} block to ensure that the \code{fieldA} of an object is updated without interference by other threads.

The figure indicates three possible program points with a number, in which execution can be suspended for the atomic block.
Further, it shows that the atomic block is known to the debugger via the \code{Atomic} tag, which it received as part of the meta data and source information.
When execution is suspended, the debugger received the source location, the activity type, and active dynamic scopes, which are depicted in the right hand side of the figure.
From the meta data and the \code{Atomic} tag associated with the source location for \circled{1}, the debugger can derive that it can offer a breakpoint that is triggered right before a transaction is started.

Setting the breakpoint sends a \code{BreakpointUpdate} message to the \SOMns interpreter, which includes the source location and the chosen breakpoint type.
Afterwards, the program can stop at the \code{atomic} keyword,
and the interpreter sends a \code{Stopped} message to the debugger.
The message says that execution is suspended at location \circled{1}, and that the current activity is a thread with a specific id.
Based on this location information, source tags, and execution information, the debugger can derive the applicable stepping operations.
Note that in this case there is no concurrent stepping applicable. 
However, all stepping operations are handled uniformly.
When the debugger determines the stepping operations for \circled{1}, the resulting set contains only the sequential stepping operations that always apply, \eg step into and step over.

When the developer chooses the step-into operation, the debugger sends a \code{Step} message to the interpreter, which includes the thread id and the chosen stepping operation.
After the interpreter performs this step, execution is suspended at \circled{2} and the debugger receives again the current location and activity.
It also receives the information that a dynamic scope for a transaction is active.
Based on this scope, it can offer extra stepping operations, \eg, to step right before or after the commit for the transaction.
When executing these stepping operations, execution would continue either to \circled{3}, or to the first statement after the \code{atomic} block.

\subsection{Detailed Description of the \Kompos Protocol}
\label{sec:protocol-details}

This section details the protocol.
While we discuss its semantics, we refrain from prescribing a specific implementation, since we consider the ideas to be applicable to wide range of concrete debugger protocols.
A concrete prototype implementation is discussed in \cref{sec:impl}.

The protocol assumes bidirectional communication between interpreter and debugger, which could be realized with remote function calls, messages on sockets, etc.

\begin{figure}
\centering
\ifpdf
\includegraphics{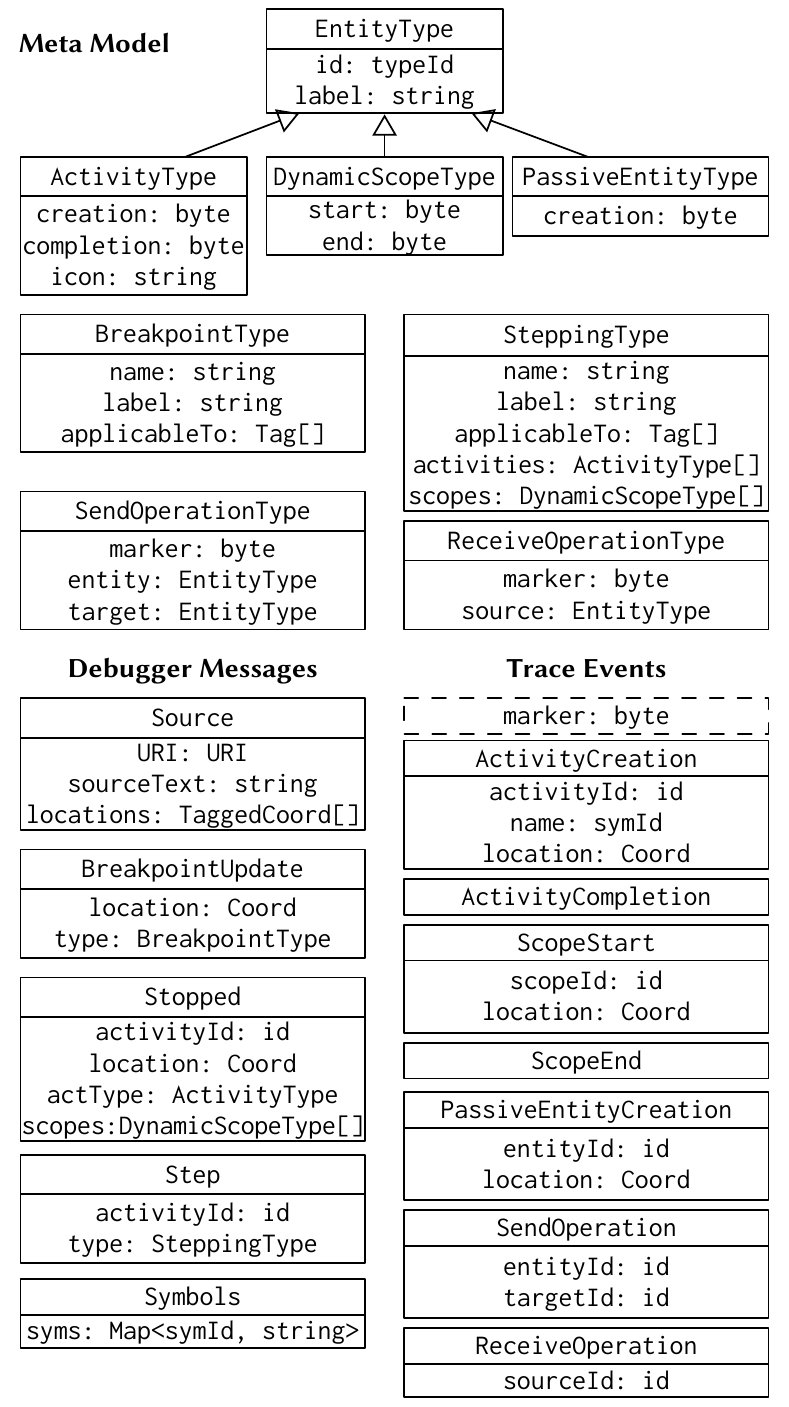}
\else
\HCode{<img src="figures/protocol-class-diagram.svg" />}
\fi
\caption{Class diagram of the main elements of the \Kompos protocol.
The meta model describes the concurrency and debugger concepts supported by the interpreter, and provides the debugger with the meta data to identify when and where breakpoints and stepping operations are applicable.
Debugger messages are used to update the debugger or interpreter.
Trace events encode an execution trace used for visualization. Trace events are prefixed with a \code{marker} byte.}
\label{fig:protocol-details}
\vspace{-10pt}
\end{figure}

\Cref{fig:protocol-details} shows an overview of the main concepts used by the \Kompos protocol.
We distinguish between (1) debugging-specific messages exchanged between debugger and interpreter, (2) trace data sent from the interpreter for visualization, and (3) a general meta-data description used to interpret the exchanged messages.
We detail each of them below.

\paragraph{Debugger Messages.}

%
%
%

For stepping and breakpoints, the \Kompos protocol uses the first four debugger messages in \cref{fig:protocol-details}.

The \code{Source} message provides the source information to the debugger.
Since the debugger is to be agnostic from specific concurrency concepts, as well as incidentally from a specific language, we provide source information explicitly.
The message includes a URI to identify the source file or resource, the source text, and a list of tagged source locations.
Source locations specify the exact coordinates, for instance based on a line number, column number, and character length.
The tags are merely opaque identifiers which identify concurrency operations, \eg, as seen with the \code{Atomic} tag in \cref{fig:stm-example}.

The \code{BreakpointUpdate} message is used to communicate breakpoints from the debugger to the interpreter.
It encodes the source location and the breakpoint type.

The \code{Stopped} message is sent from the interpreter to the debugger to indicate that either a breakpoint was hit or a stepping operation completed.
It identifies the current location and the suspended activity with id and type.
Furthermore, it includes a list of currently active dynamic scopes for this activity.
Note that the activity type and active dynamic scopes can also be determined from the trace data, but providing them explicitly simplifies the debugger implementation.

Finally, the \code{Step} message is sent from the debugger to the interpreter to instruct the latter to resume execution of a specified activity with a given stepping type.

The last message listed in \cref{fig:protocol-details}, called \code{Symbols} message, is an optimization.
It avoids sending long strings repeatedly by sending a symbol table from the interpreter to the debugger.

\paragraph{Execution Trace Data.}

%
%
%
%
%
%
%

To provide details about the execution of a concurrent program, the \Kompos protocols uses trace events that encode the program's behavior with 7 different trace entries.
We use these trace events for instance to visualize concurrent interactions.
In general, each trace event starts with a marker, which is indicated by the dashed line in \cref{fig:protocol-details}.
The relation between the concrete marker and a concurrency concept is defined via the meta data discussed in the following subsection.

An \code{ActivityCreation} event records the id of the created activity, its name, and the source location of the creation operation.
An \code{ActivityCompletion} event is merely a marker recording that an activity terminated.
The corresponding activity id can be determined from the complete trace.
A \code{ScopeStart} event records the beginning of a dynamic scope.
It records the id of a scope, which corresponds to, \eg, the message id for an actor turn.
It also includes the source location for the scope, \eg, the method invoked for a turn, or the atomic code block for a transaction.
A \code{ScopeEnd} event is also a marker that can be matched to the scope start implicitly.
A \code{PassiveEntityCreation} event records the id of the passive entity created and the source location of the operation. 
 
Interactions are recorded as \code{SendOperations} with the id of the involved passive entity, \eg, channel or message, and the target entity id, \eg, the receiving actor.
Information about the sending entity can be inferred from the trace based on the dynamic scope or current activity.
\code{ReceiveOperations} encode merely the id for the source entity which is for instance a channel or fork/join task.

\paragraph{Meta Data Description.}


The debugger messages and trace events discussed above are completely independent from concurrency models.
To distinguish different types of entities and interactions, the interpreter sends meta data to the debugger when the connection is initialized.
The meta data consists of the 8 concepts shown at the top of \cref{fig:protocol-details}.

%
%
%

There exist three types of entities:  \code{ActivityType}, \code{Dyna}\-\code{micScopeType}, and the \code{PassiveEntityType}.
\code{EntityType} defines data common to all entity types.
All entities have a label, \ie a name, and a unique id to distinguish them.
\code{ActivityType} additionally defines unique trace event marker for activity creation and completion, as well as an identifier for an icon to be used in the user interface.
\code{DynamicScopeType} defines the start and end markers for scopes, and \code{Passive}\-\code{EntityType} defines creation marker.

%
%

The \code{BreakpointType} defines the possible breakpoints.
Each type has a unique name, a label to be used in the user interface, and an applicability criterion based on source tags.
If a source location has one of the listed tags, then it supports the breakpoint.
If a breakpoint type does not specify any source tags, it applies to all source locations.

The \code{SteppingType} defines the stepping operations.
Each type has a name and label.
As for breakpoints, source tags also define whether a stepping operation is applicable.
For instance, the operation to step to the receiver of a message is only available on a message-send operation.
An additional applicability criterion is the current activity type, \eg, to enable stepping to the next turn for actors.
Similarly, stepping can be conditional to the current dynamic scope.
As such the third applicability criterion is the current scope type.  
For instance, some transaction-related stepping operations are only useful within a transaction (cf. \cref{sec:protocol-example}).

%

Finally, the meta data specifies how to interpret \code{Send}\-\code{Ope}\-\code{rations} and \code{ReceiveOperations}.
For each operation, a unique marker is defined.
For a send, the operation type specifies the entity types for the involved entity and target to identify which set of entities it belongs to.
Similarly, for a receive operation, the type of the source entity is specified.

This meta data makes it possible to handle breakpoints and stepping operations in an abstract manner.
Furthermore, it becomes possible to interpret the trace events either generically or specific to a concurrency model to visualize them.
We evaluate both aspects in \cref{sec:evaluation}.

\section{Implementation}
\label{sec:impl}


This section provides basic details on our implementation, which is used for the evaluation.
The \Kompos debugger is a TypeScript application running in a browser.
The \SOMns interpreter implements the support for the concurrency models, their breakpoints, stepping semantics, and execution tracing.
As shown in \cref{fig:architecture}, the \SOMns interpreter and the \Kompos debugger communicate via a bi-directional connection, for which we use \citeurl{Web Sockets.}{The WebSocket Protocol}{IETF}{2017-05-16}{https://tools.ietf.org/html/rfc6455}
JSON is used to encode the meta data and debugger messages.
For efficiency, the trace events are sent through a separate binary web socket.


When the \Kompos debugger connects to \SOMns, the interpreter sends the meta data to initialize the debugger.
The debugger then processes the meta data to enable efficient parsing of trace events, to initialize breakpoints, stepping operations, and visualizations.
Based on the labels provided as part of the meta data, the \Kompos debugger also interprets the meta data to enable filtering and querying data for specific concurrency models, which can be used to build tools specific to a concurrency model.


When a program executes in the \SOMns interpreter, it sends source code and source tags as part of the \code{Source} message to the debugger.
The \Kompos debugger uses this data to display the code, indicate possible locations for breakpoints, and also apply syntax highlighting based on the tags.
This approach makes the debugger completely language-agnostic.

When the program hits a breakpoint or completes a step, the debugger uses the data from the \code{Stopped} message to highlight the source location.
It also uses the meta data and information about current activity type and active dynamic scopes to select the possible stepping operations.
To obtain information on the stack trace and local variables, we use messages similar to the Visual Studio Code debugger protocol (cf. \cref{sec:dbg-protocol}).
As a result, we can also use \citeurl{Visual Studio Code as debugger for \SOMns,}{\SOMns VS Code Extension}{}{2017-05-16}{https://github.com/smarr/SOMns-vscode} for sequential debugging.
 

One challenge for the correct implementation of a concurrent debugger such as \Kompos is that interactions between the interpreter and debugger need to handle data races.
For instance, there is a race between the two web socket connections, because the order, in which messages are received between the two connections, is not guaranteed.
This can be problematic because we might hit a breakpoint for which the debugger does not yet know the corresponding activity.
We solve this in the debugger by using promises for the activities, which delays handling for the debugger message until all data is available.
Similarly, a trace event can also use a symbol id, for which the full string was not yet received via the \code{Symbols} message.
For trace events, we handle these races by waiting for all dependent data elements before a trace event can be used further.

\paragraph{General Requirements}

For an application of the \Kompos protocol to other systems, we see as a main
requirement the information about the lexical location of concurrency operations. While we
leverage Truffle's approach of annotated AST nodes, the information can also be
obtained from bytecodes or with static analysis. A larger hurdle for adoption could be the implementation of the necessary
runtime support for stepping, breakpoints, and trace events inside an existing
VM or runtime system. It might require substantial changes to ensure an efficient
implementation and provide the fine-grained stepping semantics as provided in
\SOMns. Generally, Truffle's AST-based implementation and optimizations are
convenient, but not essential. Similarly, the protocol is language and concept agnostic, and thus, can be applied to a wide range of systems.

\section{Evaluation}
\label{sec:evaluation}

This section evaluates the \Kompos protocol with respect to its ability to support breakpoints, stepping operations, and visualizations.
The goal of the evaluation is to demonstrate that the protocol is agnostic of specific concurrency abstractions and general enough to support a wide range of concurrency models.
We base the evaluation on the five aforementioned models: CEL, CSP, fork/join, STM, and threads and locks.
These models are chosen for their different concurrency characteristics, and for being the main programming models in the field.
All reported experiments are implemented as part of \citeurl{\SOMns}{\SOMns and the \Kompos Debugger Protocol}{}{2017-05-16}{https://github.com/smarr/SOMns} and the \Kompos debugger\citep{Marr:2017:KomposDemo}.



\subsection{Breakpoints}

To evaluate the flexibility of the system to represent various breakpoints, we apply the analysis results of \cref{sec:concepts} to \SOMns.
Specifically, we identify and implement \NumBreakpoints different breakpoints that can be used to pause execution based on the concurrency abstractions and their interactions.
As a general principle, we consider the point in time right before or after a concurrent operation as potentially relevant.
The goal is to allow developers to observe the effects of an operation that might interleave with other operations in the system.
The breakpoints are listed with a brief description in \cref{tab:eval-breakpoints}.

With the concepts of the \Kompos protocol presented in \cref{sec:protocol-details}, we were able to model all breakpoints solely by specifying the source tag to identify the source locations they apply to.
No specific support was required in the debugger.
The breakpoint implementation is thus completely confined to the interpreter, where the concurrency operations are implemented.
Arguably, the \Kompos protocol allows developers to define arbitrary breakpoints specific to concurrency operations, or other kind of language constructs.

\begin{table*}
\caption{Set of breakpoints implemented in \SOMns.
None of the breakpoints requires special support in the \Kompos protocol.
Instead, they are all implemented based on meta data that includes name and source tag.}
\begin{center}
\scriptsize
\begin{tabular}{l l p{8.5cm} l}
\toprule
Models &      Name               & Description          & Source Tag \\
all  & activity creation  & before an actor, process, task, or thread is created   & \code{ActivityCreation}   \\
CSP, F/J, T\&L & activity execution & before first statement of the new activity is executed & \code{ActivityCreation}   \\
   
CSP, F/J, T\&L & before join & before waiting that a process, task, or thread completes      & \code{ActivityJoin} \\
CSP, F/J, T\&L & after join  & after a process, task, or thread completed execution          & \code{ActivityJoin} \\[4pt]
   
CEL & actor message send     & before an actor message is sent                           & \code{EventualMessageSend} \\
CEL & actor message receiver & before first statement of message is processed in the receiver & \code{EventualMessageSend} \\
CEL & before async. method activation & before the first statement of a method activated by an async. msg send & \code{EventualMessageSend} \\
CEL & after  async. method activation & before returning from a method activated by an async. msg send       & \code{EventualMessageSend} \\
CEL & before promise resolution & before a promise is resolved with a value or error             & \code{PromiseCreation} \\
CEL & on promise resolution     & before the first statement of all handlers registered on promise & \code{PromiseCreation} \\[4pt]
  
CSP & before channel send    & before executing the send on a channel (set on send operation)       & \code{ChannelWrite} \\
CSP & after channel receive  & after receiving a message from a channel (set on send operation)     & \code{ChannelWrite} \\
CSP & before channel receive & before receiving a message from a channel (set on receive operation) & \code{ChannelRead}  \\
CSP & after channel  send    & after sending on a channel (set on receive operation)                & \code{ChannelRead}  \\[4pt]
  
STM & before transaction & before starting a transaction & \code{Atomic} \\
STM & before commit      & before attempting to commit changes of a transaction & \code{Atomic} \\
STM & after  commit      & after committing the transaction succeeded           & \code{Atomic} \\[4pt]

T\&L & before acquire  & before attempting to acquire a lock & \code{AcquireLock}\\
T\&L & after acquire   & after acquiring a lock              & \code{AcquireLock}\\
T\&L & before release  & before releasing a lock             & \code{ReleaseLock}\\
T\&L & after release   & after releasing a lock              & \code{ReleaseLock}\\

\bottomrule
\end{tabular}
\end{center}
\label{tab:eval-breakpoints}
\end{table*}

\subsection{Stepping Operations}

To evaluate the \Kompos protocol's support for standard and advanced stepping operations, we apply the results of the analysis in \cref{sec:concepts} and implement \NumSteppingOps different stepping operations.
Guided by the breakpoints in \cref{tab:eval-breakpoints}, we identified stepping operations that allow one to follow the execution flow between various potential points of interest.
The stepping operations are listed with a brief description in \cref{tab:eval-stepping}.

With the \Kompos protocol, we were able to model all those stepping operations and customize their applicability based on a current source location, the type of the current activity, or active dynamic scopes.
Other than these generic concepts, no support is required in the debugger.
Similar to the breakpoint support, the stepping operations are defined completely in the interpreter, where the concurrency operations are implemented.
This demonstrates that the \Kompos protocol provides the desired flexibility to define arbitrary stepping operations.

\begin{table*}
  \caption{Set of stepping operations implemented in \SOMns.
  None of these stepping operations require special support in the \Kompos protocol.
  Instead, they are realized solely based on the applicability criteria provided as part of the meta data.}
\begin{center}
\scriptsize
\begin{tabular}{l l p{7cm} l}
\toprule
Model & Name               & Description          & Criteria \\
all   & resume             & continue execution of current activity & \\
all   & pause              & pause execution of current activity    & \\
all   & stop               & terminate program                      & \\
all   & step into          & step into method call                  & \\
all   & step over          & step over method call                  & \\
all   & return             & return from method call                & \\[4pt]

CSP, F/J, T\&L & step into activity   & halt new activity before execution of the first statement &    source tag: \code{ActivityCreation} \\
CSP, F/J, T\&L & return from activity & halt activity that joins with the current one, after joining & current activity: Process, Task, Thread \\[4pt]

CEL & step to message receiver & halt activity before executing the first statement of a received message & source tag: \code{EventualMessageSend} \\
CEL & step to promise resolver   & halt activity before resolving a promise & source tag: \code{PromiseCreation} \\
CEL & step to promise resolution & halt all activities before executing the first statement of handlers registered on a promise & source tag: \code{PromiseCreation} \\
CEL & step to next turn & continue current actor's execution and stop before the first statement of the next executed message & current activity: Actor\\
CEL & return from turn to resolution & continue current actor's execution and stop before the execution of the first statement of all handlers registered on a promise that is resolved by the current turn & current activity: Actor \\[4pt]

CSP & step to channel receiver & halt activity reading from a channel to receive the sent message & source tag: \code{ChannelWrite}\\
CSP & step to channel sender   & halt activity sending to a channel & source tag: \code{ChannelRead}\\[4pt]

STM & step to next transaction  & halt activity before starting the next transaction & \\
STM & step to commit            & halt activity before committing a transaction & dynamic scope: Transaction \\
STM & step after commit         & halt activity after committing a transaction &  dynamic scope: Transaction \\[4pt]

T\&L & step to release  & halt activity before releasing a lock & dynamic scope: monitor \\
T\&L & step to next acquire & halt the next activity right after acquiring the current lock & dynamic scope: monitor \\

\bottomrule
\end{tabular}
\end{center}
\label{tab:eval-stepping}
\end{table*}

\subsection{Visualization}
Finally, we evaluate whether the data provided by the \Kompos protocol can be used for visualizations.
To this end, assess whether it is possible to built a visualization that is agnostic to the concurrency models as well as one that is specifically designed for one concurrency model.
The goal is to identify where the boundary is between concurrency-agnostic aspects and special-purpose constructs. 
We built: (1) an agnostic visualization of interactions between entities in a program, and (2) a visualization specific to the actor model which shows the execution of actor turns and their causal relationships based on the messages sent in a turn.

\subsubsection{System Interaction Visualization}
\label{sec:eval-system-view}

\begin{figure}
  \ifpdf
    \includegraphics[width=0.8\columnwidth]{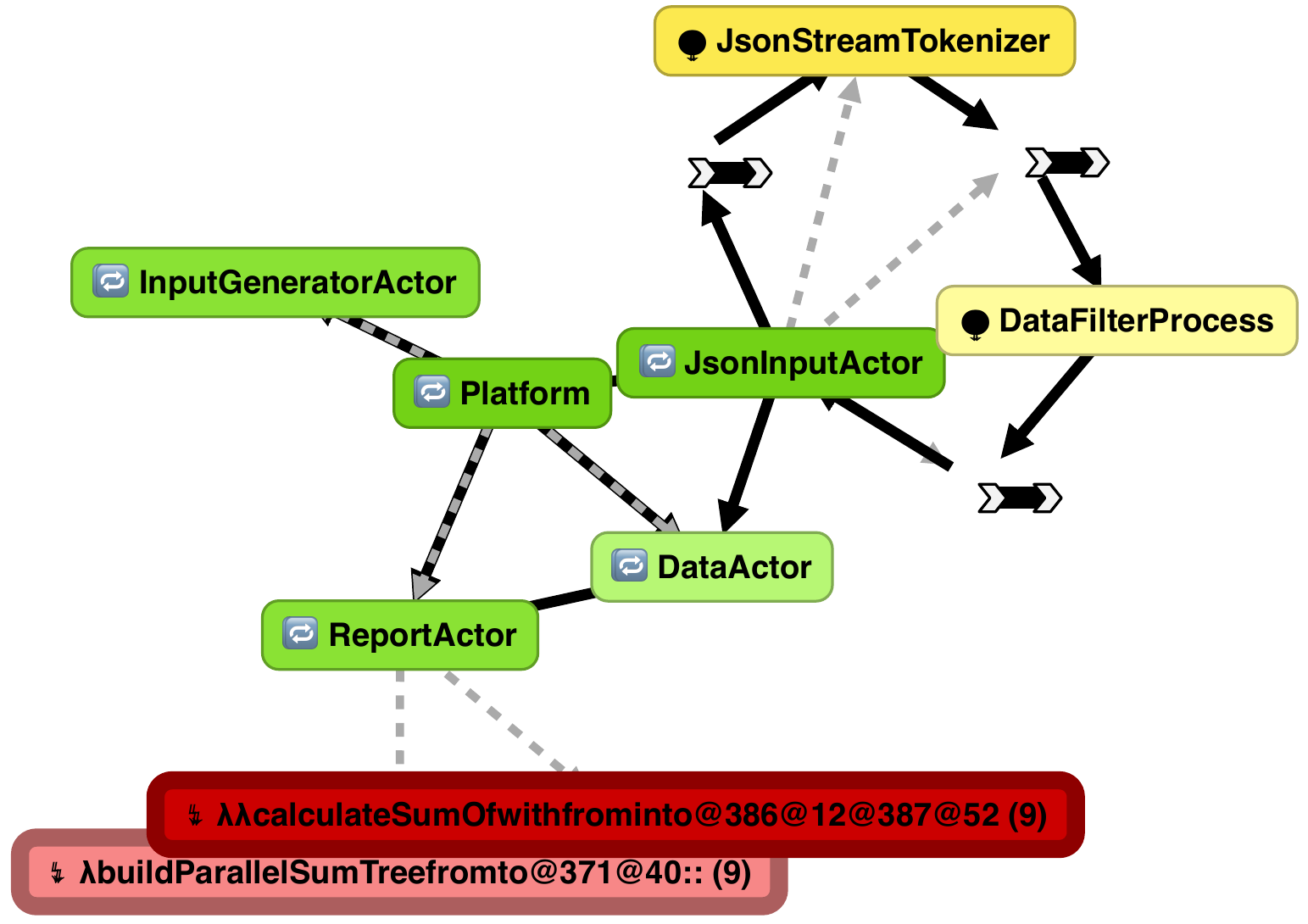}
  \else
    \HCode{<img src="figures/system-view.svg" />}
  \fi
  \caption{Screenshot of the system interaction visualization in \Kompos.
Activities are represented as rectangles.
The debugger assigns color ranges to an activity type, \eg, green for actors, yellow for processes, and red for tasks.
Color shades distinguish between activities of the same type.
Black arrows represent messages sent and gray dashed arrows indicate who created an entity. 
Black bars with two white arrows are a custom visualization for channels.
The visualization is chosen via an optional map using an entity type's label.}
  \label{fig:system-view}
\end{figure}

The system interaction visualization shows how entities communicate with each other.
\Cref{fig:system-view} shows a screenshot of the visualization.
Activities are visualized as rectangles with rounded corners.
Depending on the number of activities created from the same source location, the visualization groups the activities.
Furthermore, the debugger chooses a different color range depending on the activity type. 
On the other hand, the icon in front of an activity's name is directly specified as part of the meta data (cf. \cref{fig:protocol-details}).
An \code{ActivityType} includes a name for an icon, for which the debugger can then determine a suitable visualization.

Passive entities are visualized with custom SVG graphics.
\Cref{fig:protocol-details} shows channels as two white arrows on top of a black bar.
The visualization is generated in the debugger and matched to a \code{PassiveEntityType} based on its label.
The goal was to make these entities easier to recognize.
The design tradeoff here is between including more meta data in the protocol and leaving room for the debugger to add custom visualizations like this.
We decided that the simplest would be to have an extensible map in the debugger to select a specific visualization for entities it is aware of.

The gray-dashed arrows between entities, \ie, activities or passive entities, are determined based on their creation information.
The visualization does not show dynamic scopes.
It merely uses them to identify the connection between entities.
Actor and channel messages are not shown because we do not record their creation, but rather the specific send/receive operations.
Thus, the visualization itself is agnostic from the concurrency models, but it depends on how the data is encoded in trace events for a specific concurrency model.

Send/receive events are used for the black arrows.
Entities exchanging more messages are displayed closer together.

Overall, the system interaction visualization is independent of specific concurrency models.
It just uses the knowledge about activities, dynamic scopes, passive entities, creation operations, and send/receive operations to generate a graph representing the systems interaction.
For most aspects, the debugger independently visualizes the elements, colors, and icons considering only meta data provided by the interpreter.
However, to provide the extra bit of polish, \ie, to provide an iconic representation of channels, it includes a map of additional visualizations that matches entity type labels.
We consider this design a reasonable tradeoff that shows that small customizations are possible, but a concurrency-agnostic visualization is feasible.

\subsubsection{Actor Turn Visualization}
\label{sec:eval-actor-view}

\begin{figure}
  \ifpdf
  \includegraphics[width=0.5\columnwidth]{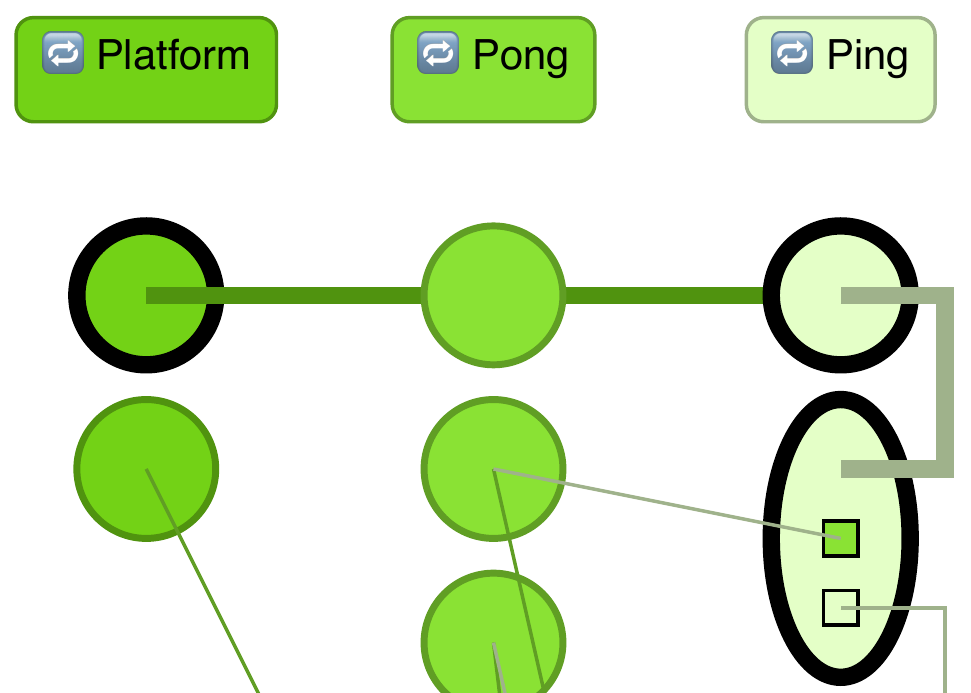}
  \else
  \HCode{<img src="figures/process-view.svg" />}
  \fi
  \caption{Screenshot of the actor turn visualization in \Kompos.
Each actor is shown on a lane with its turns indicated as circles.
Lines between turns indicate message sends.
When expanding a turn, it shows the order of the messages sent.}
  \label{fig:process-view}
\end{figure}

\Cref{fig:process-view} shows our second visualization which is inspired by the processes view in Causeway\citep{Stanley:2009}.
The goal of this visualization is to show the causality between turn executions and messages.
It visualizes each actor in the system on a lane, on which its turns are indicated as circles.
A line indicates the message that caused a turn.
When inspecting a specific turn, it unfolds into an ellipse and shows sent messages (as rectangles) in the order they were sent.
The messages connect with arrows to the turns on the receiving actor that processes them.


While this visualization is specific to the actor model, parsing and interpreting of the trace events is still done in an agnostic way.
Only after obtaining the data, the \Kompos debugger uses the meta data to determine which activities are actors, which dynamic scopes are turns, and which send operations are actor messages.

%

This visualization is specific to the notion of communicating event loops, and its implementation makes assumptions about which interactions are possible.
Nonetheless, it is based on Lamport's general \emph{happens-before relationship}\citep{Lamport:1978:TCO}, which can be applied to other concurrency models, too. 
Moreover, its implementation in \Kompos is based on the abstract notions of the protocol, and merely filters out the actor-related trace events.
Thus, it seems feasible to extent it to other concurrency models, especially if they use for instance transactions or object monitors as dynamic scopes, to indicate their causal relations.

\subsection{Conclusion}

The evaluation shows that the \Kompos protocol is abstract enough to support arbitrary breakpoints and stepping operations independent from a specific concurrency model.
Furthermore, the provided data is generic enough for tools that are agnostic of the concurrency models as shown with our system interaction view.
However, it remains possible to build tools specific to a concurrency model by interpreting the meta data, as we have shown with the actor turn view.


\section{Related Work}

This section discusses concurrent debuggers and novel IDE designs that influenced our work or that are closely related.
Debugger protocols similar to ours and their limitations have been discussed in \cref{sec:dbg-protocol}.
Generally, their support for concurrency models is minimal and they do not provide any facilities for custom breakpoint or stepping types, while the \Kompos protocol is designed for this purpose.

\subsection{Concurrent Debuggers}
\label{sec:debuggers}
%

Debuggers for concurrent and parallel systems have a long history\citep{McDowell:1989}.
This includes support for breakpoints, stepping, and visualizing of parallel systems.
However, to the best of our knowledge, so far, no debugger supports a wide range of concurrency models.

REME-D\citep{GonzalezBoix:2014} is the closest related work.
It is also an online debugger focusing on distributed communicating event-loop programs, that uses a meta-pro\-gramming API to realize breakpoints and stepping semantics.
Our actor breakpoint and stepping operations are reminiscent of REME-D's ones.
However, REME-D's API is specific to the actor model, and does not abstract from concurrency concepts as the \Kompos protocol does.

\citeurl{Erlang}{Debugger}{Ericsson AB}{2017-05-16}{http://erlang.org/doc/apps/debugger/debugger_chapter.html} and \citeurl{ScalaIDE}{Asynchronous Debugger}{ScalaIDE}{2017-05-16}{http://scala-ide.org/docs/current-user-doc/features/async-debugger/index.html} support basic debugging of actor programs with sequential stepping and breakpoints.
%
ScalaIDE also includes an option to follow a message send and stop in the receiving actor.
However, neither of them attempts to go beyond this basic debugger functionality.

\citet{Zyulkyarov:2010:DPU} introduced a debugger for a transactional system.
The focus of their work is to ensure that the STM implementation does not interfere with the debugging experience, and that stepping over or into transactions works naturally.
Furthermore, they provide mechanisms for conflict-point discovery and debug-time transactions.
Our work, however, focuses on advanced breakpoint and stepping semantics.
Their advanced debugging mechanisms would be highly interesting for \Kompos, too.

Early prototypes of the \Kompos debugger were presented by \citet{TorresLopez:2016:TAD} and \citet{Marr:2017:KomposDemo}.
However, this was only an exploration of initial ideas and did not yet include any work on the \Kompos protocol.

\subsection{Novel IDE Designs}
\label{sec:novelIDEdesigns}

Projects such as the \citeurl{Language Server Protocol,}{Language Server Protocol}{Microsoft}{2017-05-16}{https://github.com/Microsoft/language-server-protocol} which is implemented by Visual Studio Code, and Monto\citep{Keidel:2016:IPP} try to change how we think about integrated development environments (IDEs).
Instead of using the plugin approach common to Eclipse or Visual Studio, they provide support for languages by providing a common protocol to exchange information for code completion, code errors, and other common IDE services.
We consider their design an inspiration for this work. However, neither the language server protocol nor Monto support debugging at this point.

With respect to flexible debuggers, \citet{Ressia:2012:OD} explored how to bring
the abstraction level from a stack-centered view to the object level introducing
higher-level stepping operations and breakpoints.
\citet{Chis:2015:MDF} followed this line of work with a debugger framework for domain-specific debuggers.
They support domain-specific breakpoints, stepping operations, and debugger views.
For example, they have a debugger for a parser framework to step through the parsing process on the level of the parser rules instead of the parser implementation.
Similarly, they have a debugger for a complex notification system to step through the activations of the subscriptions to notifications instead of working on the basic notion of method calls and callbacks.
Instead of providing a framework for building debuggers, our work focuses on the protocol between the debugger and the interpreter.
To our understanding, our protocol supports all required elements to also support their domain-specific breakpoint and stepping operations.
However, we do not provide a framework to build custom debugger interfaces as they do.

\section{Conclusion and Future Work}

To enable better debugging tools for complex concurrent applications, we propose the \Kompos protocol, a concurrency-agnostic debugger protocol.
The protocol abstracts from specific concurrency models to support custom breakpoints, stepping operations, and visualizations, without requiring support for the specific concurrency models. 

Based on our study of shared-memory and message-passing models, the protocol represents concurrency concepts in terms of activities, dynamic scopes, and passive entities.
It uses opaque meta data to allow the debugger to determine where breakpoints or stepping operations are applicable.
The protocol also includes the notion of send and receive operations to, \eg, visualize concurrent interactions.

To evaluate the protocol, we implemented it in the \Kompos debugger and \SOMns.
\SOMns supports the five major concurrency models: threads and locks, communicating event loops, communicating sequential processes, fork/join parallelism, and software transactional memory.
We implemented \NumBreakpoints breakpoints and \NumSteppingOps stepping operations in \SOMns for these models, without requiring any modifications to the debugger, which shows that the protocol is concurrency agnostic.
We also implemented two visualizations.
The first one shows the concurrent interactions independently of the concurrency models.
The second one shows causalities between actor turns, messages, and their ordering, which is specific to the communicating event-loop model.
This demonstrates that the protocol is flexible enough to enable advanced debugging tools that are concurrency agnostic, while it remains possible to build tooling specific to a concurrency model.

Based on this work, existing debugger protocols could be extended to provide advanced debugging support for concurrent programming, without requiring support for specific concurrency models.
This provides a foundation for better tooling and debuggers for complex concurrent systems that combine concurrency models.

For future work, we would like to study how to enable arbitrary libraries to benefit from such a generic protocol.
The challenge here is to expose the relevant data about concepts and their relation to library methods to the interpreter so that it can be communicated to the debugger.
Especially in dynamic languages, it needs to be able to expose this information at runtime.

Further work is also required to make the visualization scalable to large applications.
We need to find ways to explore complex systems and focus on relevant details,
and we need to investigate ways to provide the relevant data efficiently.
Future work also needs to study how to effectively expose the large number of concurrency-specific debugger features to users, and whether they help to debug concurrent applications more effectively.



\begin{acks}                            

We would like to thank Sander Lenaerts for the implementation of the actor turn view, and Manuel Rigger and Richard Roberts for comments on an early draft.
Stefan Marr and Dominik Aumayr were funded by a grant of the \grantsponsor{FWF}{Austrian Science Fund}{http://www.fwf.ac.at/} (FWF), project number \grantnum{http://www.fwf.ac.at/}{}{I2491-N31}.
Carmen Torres Lopez was funded by a grant of the \grantsponsor{FWO}{Research Foundation Flanders}{http://www.fwo.be/} (FWO), project number \grantnum{http://www.fwo.be/}{}{G004816N}.

%
\end{acks}

~~\\[10pt]

\newcommand{\showDOI}[1]{\unskip}
\newcommand{\showURL}[1]{\unskip}
\bibliography{references,bibsonomy}


\begin{thebibliography}{00}


\ifx \showCODEN    \undefined \def \showCODEN     #1{\unskip}     \fi
\ifx \showDOI      \undefined \def \showDOI       #1{#1}\fi
\ifx \showISBNx    \undefined \def \showISBNx     #1{\unskip}     \fi
\ifx \showISBNxiii \undefined \def \showISBNxiii  #1{\unskip}     \fi
\ifx \showISSN     \undefined \def \showISSN      #1{\unskip}     \fi
\ifx \showLCCN     \undefined \def \showLCCN      #1{\unskip}     \fi
\ifx \shownote     \undefined \def \shownote      #1{#1}          \fi
\ifx \showarticletitle \undefined \def \showarticletitle #1{#1}   \fi
\ifx \showURL      \undefined \def \showURL       {\relax}        \fi
\providecommand\bibfield[2]{#2}
\providecommand\bibinfo[2]{#2}
\providecommand\natexlab[1]{#1}
\providecommand\showeprint[2][]{arXiv:#2}

\bibitem[\protect\citeauthoryear{Almasi and Gottlieb}{Almasi and
  Gottlieb}{1994}]%
        {Almasi:1989:HPC}
\bibfield{author}{\bibinfo{person}{George~S. Almasi} {and}
  \bibinfo{person}{Allan Gottlieb}.} \bibinfo{year}{1994}\natexlab{}.
\newblock \bibinfo{booktitle}{{\em Highly Parallel Computing\/}
  (\bibinfo{edition}{2nd} ed.)}.
\newblock \bibinfo{publisher}{Benjamin-Cummings Publishing Co., Inc.}
\newblock
\showISBNx{0805304436 9780805304435}


\bibitem[\protect\citeauthoryear{Blumofe, Joerg, Kuszmaul, Leiserson, Randall,
  and Zhou}{Blumofe et~al\mbox{.}}{1995}]%
        {Cilk}
\bibfield{author}{\bibinfo{person}{Robert~D. Blumofe},
  \bibinfo{person}{Christopher~F. Joerg}, \bibinfo{person}{Bradley~C.
  Kuszmaul}, \bibinfo{person}{Charles~E. Leiserson}, \bibinfo{person}{Keith~H.
  Randall}, {and} \bibinfo{person}{Yuli Zhou}.}
  \bibinfo{year}{1995}\natexlab{}.
\newblock \showarticletitle{Cilk: An Efficient Multithreaded Runtime System}.
  In \bibinfo{booktitle}{{\em Proc. of PPoPP}}, Vol.~\bibinfo{volume}{30}.
  \bibinfo{publisher}{ACM}.
\newblock
\showISSN{0362-1340}
\showDOI{%
\url{https://doi.org/10.1145/209937.209958}}


\bibitem[\protect\citeauthoryear{Bracha, von~der Ahé, Bykov, Kashai, Maddox,
  and Miranda}{Bracha et~al\mbox{.}}{2010}]%
        {Bracha:10:NS}
\bibfield{author}{\bibinfo{person}{Gilad Bracha}, \bibinfo{person}{Peter
  von~der Ahé}, \bibinfo{person}{Vassili Bykov}, \bibinfo{person}{Yaron
  Kashai}, \bibinfo{person}{William Maddox}, {and} \bibinfo{person}{Eliot
  Miranda}.} \bibinfo{year}{2010}\natexlab{}.
\newblock \showarticletitle{{Modules as Objects in Newspeak}}.
\newblock In \bibinfo{booktitle}{{\em Proc. of ECOOP}}. \bibinfo{series}{LNCS},
  Vol.~\bibinfo{volume}{6183}. \bibinfo{publisher}{Springer},
  \bibinfo{pages}{405--428}.
\newblock
\showISBNx{978-3-642-14106-5}
\showDOI{%
\url{https://doi.org/10.1007/978-3-642-14107-2_20}}


\bibitem[\protect\citeauthoryear{Chiş, Denker, Gîrba, and Nierstrasz}{Chiş
  et~al\mbox{.}}{2015}]%
        {Chis:2015:MDF}
\bibfield{author}{\bibinfo{person}{Andrei Chiş}, \bibinfo{person}{Marcus
  Denker}, \bibinfo{person}{Tudor Gîrba}, {and} \bibinfo{person}{Oscar
  Nierstrasz}.} \bibinfo{year}{2015}\natexlab{}.
\newblock \showarticletitle{Practical domain-specific debuggers using the
  Moldable Debugger framework}.
\newblock \bibinfo{journal}{{\em Computer Languages, Systems \& Structures\/}}
  \bibinfo{volume}{44, Part A} (\bibinfo{year}{2015}),
  \bibinfo{pages}{89--113}.
\newblock
\showISSN{1477-8424}
\showDOI{%
\url{https://doi.org/10.1016/j.cl.2015.08.005}}


\bibitem[\protect\citeauthoryear{De~Koster, Van~Cutsem, and
  De~Meuter}{De~Koster et~al\mbox{.}}{2016}]%
        {DeKoster:2016:YAT}
\bibfield{author}{\bibinfo{person}{Joeri De~Koster}, \bibinfo{person}{Tom
  Van~Cutsem}, {and} \bibinfo{person}{Wolfgang De~Meuter}.}
  \bibinfo{year}{2016}\natexlab{}.
\newblock \showarticletitle{43 Years of Actors: A Taxonomy of Actor Models and
  Their Key Properties}. In \bibinfo{booktitle}{{\em Proc. of AGERE!'16}}.
  \bibinfo{publisher}{ACM}, \bibinfo{pages}{31--40}.
\newblock
\showISBNx{978-1-4503-4639-9}
\showDOI{%
\url{https://doi.org/10.1145/3001886.3001890}}


\bibitem[\protect\citeauthoryear{Gonzalez~Boix, Noguera, and
  De~Meuter}{Gonzalez~Boix et~al\mbox{.}}{2014}]%
        {GonzalezBoix:2014}
\bibfield{author}{\bibinfo{person}{Elisa Gonzalez~Boix},
  \bibinfo{person}{Carlos Noguera}, {and} \bibinfo{person}{Wolfgang
  De~Meuter}.} \bibinfo{year}{2014}\natexlab{}.
\newblock \showarticletitle{Distributed debugging for mobile networks}.
\newblock \bibinfo{journal}{{\em Systems and Software\/}}  \bibinfo{volume}{90}
  (\bibinfo{year}{2014}).
\newblock


\bibitem[\protect\citeauthoryear{Harris, Marlow, Peyton-Jones, and
  Herlihy}{Harris et~al\mbox{.}}{2005}]%
        {Harris:2005}
\bibfield{author}{\bibinfo{person}{Tim Harris}, \bibinfo{person}{Simon Marlow},
  \bibinfo{person}{Simon Peyton-Jones}, {and} \bibinfo{person}{Maurice
  Herlihy}.} \bibinfo{year}{2005}\natexlab{}.
\newblock \showarticletitle{Composable Memory Transactions}. In
  \bibinfo{booktitle}{{\em Proc. of PPoPP'05}}. \bibinfo{publisher}{ACM},
  \bibinfo{pages}{48--60}.
\newblock
\showISBNx{1-59593-080-9}
\showDOI{%
\url{https://doi.org/10.1145/1065944.1065952}}


\bibitem[\protect\citeauthoryear{Hoare}{Hoare}{1978}]%
        {CSP}
\bibfield{author}{\bibinfo{person}{C.~A.~R. Hoare}.}
  \bibinfo{year}{1978}\natexlab{}.
\newblock \showarticletitle{Communicating Sequential Processes}.
\newblock \bibinfo{journal}{{\em Commun. ACM\/}} \bibinfo{volume}{21},
  \bibinfo{number}{8} (\bibinfo{year}{1978}), \bibinfo{pages}{666--677}.
\newblock
\showISSN{0001-0782}
\showDOI{%
\url{https://doi.org/10.1145/359576.359585}}


\bibitem[\protect\citeauthoryear{Keidel, Pfeiffer, and Erdweg}{Keidel
  et~al\mbox{.}}{2016}]%
        {Keidel:2016:IPP}
\bibfield{author}{\bibinfo{person}{Sven Keidel}, \bibinfo{person}{Wulf
  Pfeiffer}, {and} \bibinfo{person}{Sebastian Erdweg}.}
  \bibinfo{year}{2016}\natexlab{}.
\newblock \showarticletitle{{The IDE Portability Problem and Its Solution in
  Monto}}. In \bibinfo{booktitle}{{\em Proc. of SLE'16}}.
  \bibinfo{publisher}{ACM}, \bibinfo{pages}{152--162}.
\newblock
\showISBNx{978-1-4503-4447-0}
\showDOI{%
\url{https://doi.org/10.1145/2997364.2997368}}


\bibitem[\protect\citeauthoryear{Lamport}{Lamport}{1978}]%
        {Lamport:1978:TCO}
\bibfield{author}{\bibinfo{person}{Leslie Lamport}.}
  \bibinfo{year}{1978}\natexlab{}.
\newblock \showarticletitle{{Time, Clocks, and the Ordering of Events in a
  Distributed System}}.
\newblock \bibinfo{journal}{{\em Commun. ACM\/}} \bibinfo{volume}{21},
  \bibinfo{number}{7} (\bibinfo{date}{July} \bibinfo{year}{1978}),
  \bibinfo{pages}{558--565}.
\newblock
\showISSN{0001-0782}
\showDOI{%
\url{https://doi.org/10.1145/359545.359563}}


\bibitem[\protect\citeauthoryear{Marr, Torres~Lopez, Aumayr, Gonzalez~Boix, and
  M\"{o}ssenb\"{o}ck}{Marr et~al\mbox{.}}{2017}]%
        {Marr:2017:KomposDemo}
\bibfield{author}{\bibinfo{person}{Stefan Marr}, \bibinfo{person}{Carmen
  Torres~Lopez}, \bibinfo{person}{Dominik Aumayr}, \bibinfo{person}{Elisa
  Gonzalez~Boix}, {and} \bibinfo{person}{Hanspeter M\"{o}ssenb\"{o}ck}.}
  \bibinfo{year}{2017}\natexlab{}.
\newblock \bibinfo{title}{{K\'ompos: A Platform for Debugging Complex
  Concurrent Applications}}.
\newblock   (\bibinfo{date}{2 April} \bibinfo{year}{2017}),
  \bibinfo{numpages}{2}~pages.
\newblock


\bibitem[\protect\citeauthoryear{McDowell and Helmbold}{McDowell and
  Helmbold}{1989}]%
        {McDowell:1989}
\bibfield{author}{\bibinfo{person}{Charles~E. McDowell} {and}
  \bibinfo{person}{David~P. Helmbold}.} \bibinfo{year}{1989}\natexlab{}.
\newblock \showarticletitle{Debugging Concurrent Programs}.
\newblock \bibinfo{journal}{{\em ACM Comput. Surv.\/}} \bibinfo{volume}{21},
  \bibinfo{number}{4} (\bibinfo{date}{Dec.} \bibinfo{year}{1989}),
  \bibinfo{pages}{593--622}.
\newblock
\showISSN{0360-0300}
\showDOI{%
\url{https://doi.org/10.1145/76894.76897}}


\bibitem[\protect\citeauthoryear{Miller, Tribble, and Shapiro}{Miller
  et~al\mbox{.}}{2005}]%
        {ELangActors}
\bibfield{author}{\bibinfo{person}{Mark~S. Miller}, \bibinfo{person}{E.~Dean
  Tribble}, {and} \bibinfo{person}{Jonathan Shapiro}.}
  \bibinfo{year}{2005}\natexlab{}.
\newblock \showarticletitle{Concurrency Among Strangers: Programming in E as
  Plan Coordination}. In \bibinfo{booktitle}{{\em Symposium on Trustworthy
  Global Computing}} {\em (\bibinfo{series}{LNCS})},
  \bibfield{editor}{\bibinfo{person}{R.~De Nicola} {and}
  \bibinfo{person}{D.~Sangiorgi}} (Eds.), Vol.~\bibinfo{volume}{3705}.
  \bibinfo{publisher}{Springer}, \bibinfo{pages}{195--229}.
\newblock
\showDOI{%
\url{https://doi.org/10.1007/11580850_12}}


\bibitem[\protect\citeauthoryear{Ressia, Bergel, and Nierstrasz}{Ressia
  et~al\mbox{.}}{2012}]%
        {Ressia:2012:OD}
\bibfield{author}{\bibinfo{person}{Jorge Ressia}, \bibinfo{person}{Alexandre
  Bergel}, {and} \bibinfo{person}{Oscar Nierstrasz}.}
  \bibinfo{year}{2012}\natexlab{}.
\newblock \showarticletitle{{Object-Centric Debugging}}. In
  \bibinfo{booktitle}{{\em Proceedings of the 34th International Conference on
  Software Engineering}} {\em (\bibinfo{series}{ICSE '12})}.
  \bibinfo{publisher}{IEEE Press}, \bibinfo{pages}{485--495}.
\newblock
\showISBNx{978-1-4673-1067-3}
\showISSN{0270-5257}
\showDOI{%
\url{https://doi.org/10.1109/ICSE.2012.6227167}}


\bibitem[\protect\citeauthoryear{Seaton, Van De~Vanter, and Haupt}{Seaton
  et~al\mbox{.}}{2014}]%
        {Seaton:2014:DFS}
\bibfield{author}{\bibinfo{person}{Chris Seaton}, \bibinfo{person}{Michael~L.
  Van De~Vanter}, {and} \bibinfo{person}{Michael Haupt}.}
  \bibinfo{year}{2014}\natexlab{}.
\newblock \showarticletitle{Debugging at Full Speed}. In
  \bibinfo{booktitle}{{\em Proc. of DYLA'14}}. \bibinfo{publisher}{ACM},
  Article \bibinfo{articleno}{2}, \bibinfo{numpages}{13}~pages.
\newblock
\showISBNx{978-1-4503-2916-3}
\showDOI{%
\url{https://doi.org/10.1145/2617548.2617550}}


\bibitem[\protect\citeauthoryear{Stanley, Close, and Miller}{Stanley
  et~al\mbox{.}}{2009}]%
        {Stanley:2009}
\bibfield{author}{\bibinfo{person}{Terry Stanley}, \bibinfo{person}{Tyler
  Close}, {and} \bibinfo{person}{Mark Miller}.}
  \bibinfo{year}{2009}\natexlab{}.
\newblock \bibinfo{booktitle}{{\em {Causeway: A message-oriented distributed
  debugger}}}.
\newblock \bibinfo{type}{{T}echnical {R}eport}. \bibinfo{institution}{HP Labs}.
  \bibinfo{pages}{1--15} pages.
\newblock
\newblock
\shownote{HP Labs tech report HPL-2009-78.}


\bibitem[\protect\citeauthoryear{Tasharofi, Dinges, and Johnson}{Tasharofi
  et~al\mbox{.}}{2013}]%
        {CastagnaECOOP2013}
\bibfield{author}{\bibinfo{person}{Samira Tasharofi}, \bibinfo{person}{Peter
  Dinges}, {and} \bibinfo{person}{Ralph~E. Johnson}.}
  \bibinfo{year}{2013}\natexlab{}.
\newblock \showarticletitle{Why Do Scala Developers Mix the Actor Model with
  other Concurrency Models?}. In \bibinfo{booktitle}{{\em Proc. of ECOOP}} {\em
  (\bibinfo{series}{LNCS})}, Vol.~\bibinfo{volume}{7920}.
  \bibinfo{publisher}{Springer}, \bibinfo{pages}{302--326}.
\newblock
\showISBNx{978-3-642-39037-1}
\showDOI{%
\url{https://doi.org/10.1007/978-3-642-39038-8_13}}


\bibitem[\protect\citeauthoryear{Torres~Lopez, Marr, Mössenböck, and
  Gonzalez~Boix}{Torres~Lopez et~al\mbox{.}}{2016}]%
        {TorresLopez:2016:TAD}
\bibfield{author}{\bibinfo{person}{Carmen Torres~Lopez},
  \bibinfo{person}{Stefan Marr}, \bibinfo{person}{Hanspeter Mössenböck},
  {and} \bibinfo{person}{Elisa Gonzalez~Boix}.}
  \bibinfo{year}{2016}\natexlab{}.
\newblock \bibinfo{title}{{Towards Advanced Debugging Support for Actor
  Languages: Studying Concurrency Bugs in Actor-based Programs}}.
\newblock   (\bibinfo{date}{30 October} \bibinfo{year}{2016}).
\newblock
\newblock
\shownote{Presentation, AGERE! '16.}


\bibitem[\protect\citeauthoryear{Van~Cutsem, Gonzalez~Boix, Scholliers,
  Lombide~Carreton, Harnie, Pinte, and De~Meuter}{Van~Cutsem
  et~al\mbox{.}}{2014}]%
        {VanCutsem2014112}
\bibfield{author}{\bibinfo{person}{Tom Van~Cutsem}, \bibinfo{person}{Elisa
  Gonzalez~Boix}, \bibinfo{person}{Christophe Scholliers},
  \bibinfo{person}{Andoni Lombide~Carreton}, \bibinfo{person}{Dries Harnie},
  \bibinfo{person}{Kevin Pinte}, {and} \bibinfo{person}{Wolfgang De~Meuter}.}
  \bibinfo{year}{2014}\natexlab{}.
\newblock \showarticletitle{AmbientTalk: programming responsive mobile
  peer-to-peer applications with actors}.
\newblock \bibinfo{journal}{{\em Com. Lan., Sys. \& Struct.\/}}
  \bibinfo{volume}{40}, \bibinfo{number}{3--4} (\bibinfo{year}{2014}),
  \bibinfo{pages}{112--136}.
\newblock
\showISSN{1477-8424}
\showDOI{%
\url{https://doi.org/10.1016/j.cl.2014.05.002}}


\bibitem[\protect\citeauthoryear{Van De~Vanter}{Van De~Vanter}{2017}]%
        {MVDV:2017:MoreVMs}
\bibfield{author}{\bibinfo{person}{Michael Van De~Vanter}.}
  \bibinfo{year}{2017}\natexlab{}.
\newblock \bibinfo{title}{{Building Flexible, Low-Overhead Tooling Support into
  a High-Performance Polyglot VM: Extended Abstract}}.
\newblock   (\bibinfo{date}{2 April} \bibinfo{year}{2017}),
  \bibinfo{numpages}{3}~pages.
\newblock
\showURL{%
\url{http://vandevanter.net/mlvdv/publications/mlvdv-morevms-2017.pdf}}
\newblock
\shownote{Presentation, MoreVMs'17.}


\bibitem[\protect\citeauthoryear{Van De~Vanter}{Van De~Vanter}{2015}]%
        {VanDeVanter:2015:BDO}
\bibfield{author}{\bibinfo{person}{Michael~L. Van De~Vanter}.}
  \bibinfo{year}{2015}\natexlab{}.
\newblock \showarticletitle{{Building Debuggers and Other Tools: We Can "Have
  It All"}}. In \bibinfo{booktitle}{{\em Proc. of ICOOOLPS'15}}.
  \bibinfo{publisher}{ACM}, Article \bibinfo{articleno}{2},
  \bibinfo{numpages}{3}~pages.
\newblock
\showISBNx{978-1-4503-3657-4}
\showDOI{%
\url{https://doi.org/10.1145/2843915.2843917}}


\bibitem[\protect\citeauthoryear{Würthinger, Wöß, Stadler, Duboscq, Simon,
  and Wimmer}{Würthinger et~al\mbox{.}}{2012}]%
        {Wurthinger:2012:SelfOptAST}
\bibfield{author}{\bibinfo{person}{Thomas Würthinger},
  \bibinfo{person}{Andreas Wöß}, \bibinfo{person}{Lukas Stadler},
  \bibinfo{person}{Gilles Duboscq}, \bibinfo{person}{Doug Simon}, {and}
  \bibinfo{person}{Christian Wimmer}.} \bibinfo{year}{2012}\natexlab{}.
\newblock \showarticletitle{Self-Optimizing AST Interpreters}. In
  \bibinfo{booktitle}{{\em Proc. of DLS'12}}. \bibinfo{pages}{73--82}.
\newblock
\showISBNx{978-1-4503-1564-7}
\showDOI{%
\url{https://doi.org/10.1145/2384577.2384587}}


\bibitem[\protect\citeauthoryear{Zyulkyarov, Harris, Unsal, Cristal, and
  Valero}{Zyulkyarov et~al\mbox{.}}{2010}]%
        {Zyulkyarov:2010:DPU}
\bibfield{author}{\bibinfo{person}{Ferad Zyulkyarov}, \bibinfo{person}{Tim
  Harris}, \bibinfo{person}{Osman~S. Unsal}, \bibinfo{person}{Adr\'{\i}an
  Cristal}, {and} \bibinfo{person}{Mateo Valero}.}
  \bibinfo{year}{2010}\natexlab{}.
\newblock \showarticletitle{Debugging Programs That Use Atomic Blocks and
  Transactional Memory}. In \bibinfo{booktitle}{{\em Proc. of PPoPP'10}}.
  \bibinfo{publisher}{ACM}, \bibinfo{pages}{57--66}.
\newblock
\showISBNx{978-1-60558-877-3}
\showDOI{%
\url{https://doi.org/10.1145/1693453.1693463}}


\end{thebibliography}

%

\end{document}